\documentclass[sigconf,natbib]{acmart}

\usepackage{bm}
\usepackage{url}
\usepackage{makecell}
\usepackage{amsmath}
\usepackage{algorithmic}
\usepackage{graphicx}
\usepackage{textcomp}
\usepackage{xcolor}
\usepackage{bbding}
\usepackage{subfigure}
\usepackage{enumitem}
\usepackage{multirow}

\newcommand{\Set}[1]{\mathcal{#1}}
\newcommand{\Mat}[1]{\bm{#1}}
\newcommand{\Vector}[1]{\bm{#1}}

\AtBeginDocument{%
  \providecommand\BibTeX{{%
    \normalfont B\kern-0.5em{\scshape i\kern-0.25em b}\kern-0.8em\TeX}}}

\setcopyright{acmcopyright}
\copyrightyear{2018}
\acmYear{2018}
\acmDOI{XXXXXXX.XXXXXXX}

\acmConference[Conference acronym 'XX]{Make sure to enter the correct
  conference title from your rights confirmation emai}{June 03--05,
  2018}{Woodstock, NY}
\acmPrice{15.00}
\acmISBN{978-1-4503-XXXX-X/18/06}

\begin{document}

\title{Estimating Difficulty Levels of Programming Problems with Pre-trained Models}

\author{Zhiyuan Wang}
\affiliation{%
  \institution{East China Normal University}
  \streetaddress{1 Th{\o}rv{\"a}ld Circle}
  \city{Shanghai}
  \country{China}}
\email{51205901111@stu.ecnu.edu.cn}

\author{Wei	Zhang}
\affiliation{%
  \institution{East China Normal University}
  \streetaddress{1 Th{\o}rv{\"a}ld Circle}
  \city{Shanghai}
  \country{China}}
\email{zhangwei.thu2011@gmail.com}

\author{Jun	Wang}
\affiliation{%
  \institution{East China Normal University}
  \streetaddress{1 Th{\o}rv{\"a}ld Circle}
  \city{Shanghai}
  \country{China}}
\email{wangjun@gmail.com	}

\renewcommand{\shortauthors}{Trovato and Tobin, et al.}

\begin{abstract}
As the demand for programming skills grows across industries and academia, students often turn to Programming Online Judge (POJ) platforms for coding practice and competition.
The difficulty level of each programming problem serves as an essential reference for guiding students' adaptive learning. 
However, current methods of determining difficulty levels either require extensive expert annotations or take a long time to accumulate enough student solutions for each problem.
To address this issue, we formulate the problem of automatic difficulty level estimation of each programming problem, given its textual description and a solution example of code. 
For tackling this problem, we propose to couple two pre-trained models, one for text modality and the other for code modality, into a unified model. 
We build two POJ datasets for the task and the results demonstrate the effectiveness of the proposed approach and the contributions of both modalities.

\end{abstract}



\maketitle

\section{Introduction}
\begin{figure}[!t]
  \centering
  \includegraphics[width=.95\linewidth]{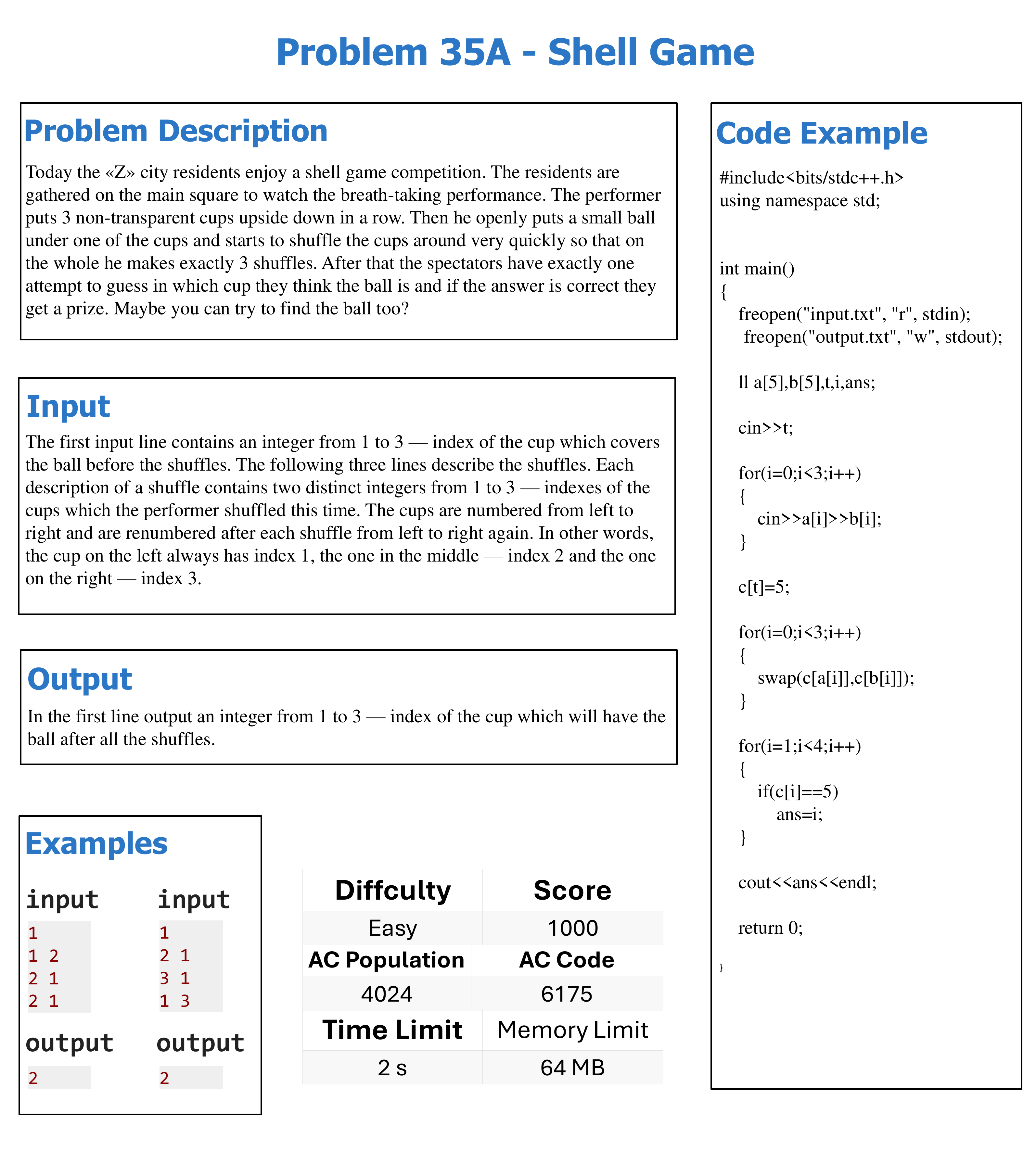}
  \vspace{-.5em}
  \caption{A real case of a programming problem and an example of code to solute the problem.}
  \label{fig:real-case}
\end{figure}

In the information era, programming skills have become increasingly crucial across various fields and industries, extending beyond just computer science and IT companies.
In response to this trend, many students with diverse academic backgrounds regularly participate in programming practice and competitions on Programming Online Judge (POJ) platforms, including but not limited to Codeforces and Leetcode.
These platforms offer an extensive range of programming problems, each accompanied by a problem statement containing a description, input and output specifications, and other relevant requirements for program implementation.
The left part of Figure~\ref{fig:real-case} illustrates an actual programming problem from one of these platforms.
The platforms are equipped with automated systems for evaluating the accuracy of code submitted by the students. The feedback provided by these platforms helps the students to revise their solutions, if necessary.

The selection of programming problems for practice is critical for effective adaptive learning on online judge platforms due to their vast repository of problems.
Previous research~\cite{Brusilovsky92} has emphasized the importance of considering the difficulty level of programming questions as a key reference for guiding the problem selection process. 
Typically, students prefer to learn questions in an order from easier to more difficult. Therefore, providing the difficulty level for each problem is essential for effective learning on online judge platforms and can facilitate downstream tasks, such as programming problem recommendation~\cite{vesin2021adaptive}.
Currently, there are two main manners to determine the difficulty: expert annotations and the correctness statistics of student solutions.
However, the former manner involves a significant amount of manual labor costs and leads to subjective assessments of difficulty, while the latter requires waiting for a sufficient number of student solutions to accumulate in order to ensure reliable statistics, although it is more objective and accurate.

In the existing literature, there has been limited exploration of automatic methods for assessing the difficulty of programming problems. Although a few studies~\cite{intisar2018cluster,EffenbergerCP19} have attempted to develop such methods, they have relied on accumulating a sufficient number of student solutions
Other relevant studies~\cite{zhao2018automatically,Mao0MPBC21,ckt4codeDKT,ck5pst} also learn from programming problems, but they focus on predicting whether a specific student could provide correct solutions to the problems.
In this paper, we formulate the task, i.e., difficulty level prediction of programming problems, to tackle the issues of the existing manners.
This task represents an innovative, multi-modal understanding problem that involves both the text modality (i.e., understanding the problem statement) and the code modality (i.e., learning from example solutions provided by programming experts), as shown in Figure~\ref{fig:real-case}.
By solving this task, it is possible to obtain a more objective assessment of problem difficulty without relying on the accumulation of student solutions, which has the potential to yield significant contributions to the field of programming education.

To enhance the modeling of both text and code modalities, we propose an approach, named C-BERT.
It leverages the power of large pre-trained models.
Specifically, BERT~\cite{devlin2018bert} is utilized to model the problem statement, while CodeBERT~\cite{fengcodebert} is used to model the example solution of codes.
As the problem statement and the example code solution are deeply relevant to each other, C-BERT further capture the interactions of the two modalities by associating their representations from each pre-trained model.
Specifically, we set two CLS tokens inside each BERT and CodeBERT, and use one CLS token as intra-modal representations and another for cross-modal representations.
Finally, the representations from the two pre-trained models are combined for difficulty level estimation.
We conduct experiments on the real datasets collected from Codeforces and CodeChef.
The results show the effectiveness of the proposed model and validate the benefits of considering both text and code modalities.


\section{Task Formulation}
Denote by $\Set{D}$ a set of POJ problems.
We suppose a given POJ problem $d\in \Set{D}$ to be composed of problem statement $w$, code example $c$ (as a solution), and explicit features $r$ (e.g., operational requirement).
The problem statement involves a sequence of words, i.e., $w=\{w_1,\cdots,w_n\}$.
The code example contains a sequence of tokens, i.e, $c=\{c_1,\cdots,c_m\}$.
And the explicit features are represented as a feature vector $\Vector{x}_r$ (shown in Section~\ref{subsec:pre-train}).
Based on these notations, the goal of this task is to learn a difficulty level estimation function $f:(w,c,r)\rightarrow y$, where $y$ denotes the difficulty level.

\section{The Computational Approach}
\begin{figure}[!t]
  \centering
  \includegraphics[width=\linewidth]{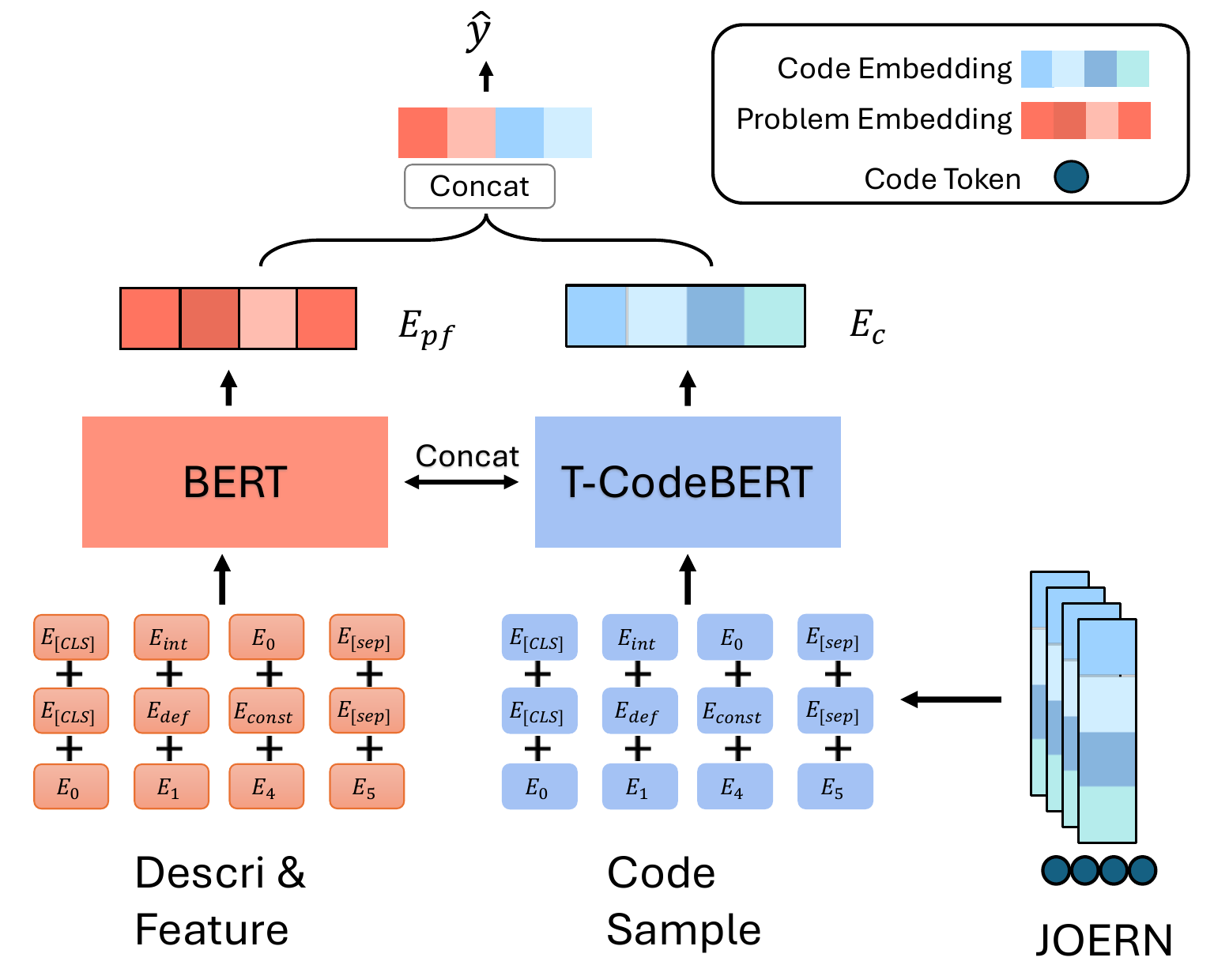}
  \vspace{-.5em}
  \caption{Architecture of the proposed approach.}
  \label{fig:model}
\end{figure}

The overview of the model architecture is shown in Figure~\ref{fig:model}.
Basically speaking, we utilize BERT to model the text modality and CodeBERT to model the code modality.
To enhance the mutual interaction of the two parts, we propose to feed the CLS embedding of each part to its counterpart, which is simple but demonstrated to be effective.
On the top layer of the model, we concatenate the text representation, the code representation, and the explicit features to conduct difficulty level estimation.
In what follows, we detail the approach.

\subsection{Basic Code and Text Representations}
For the code modeling part, we use CodeBERT and additionally consider the type of each token.
This is because tokens in the source code have different functions (e.g., variables, constants) and tokens with the same type might exhibit some similarities.
However, the original CodeBERT does not use this information.
As such, we use the code analysis tool JOERN\footnote{https://github.com/joernio/joern}
to process the source code and generate the token type information.

The proposed model C-BERT combines the token representations, the type representations, and the position representations to build the model input.
For the given programming problem, the token representation matrix, the type representation, and the position representation matrix are defined as $\Mat{E}^1_{token}$, $\Mat{E}^1_{type}$, and $\Mat{E}^1_{pos}$, respectively.
The computational formulas for CodeBERT are defined as follows:
\begin{equation}\label{eq:emb1}
\Mat{E}^1 = \Mat{E}^1_{token}+\Mat{E}^1_{type}+\Mat{E}^1_{pos}\,,
\end{equation}
\begin{equation}\label{eq:codebert}
\Mat{H}^1, \Vector{h}^1_{CLS} = \mathrm{CodeBERT}(\Mat{E}^1)\,,
\end{equation}
where $\Mat{H}^1$ is the hidden token representations and $\Vector{h}^1_{CLS}$ is the CLS embedding obtained by CodeBERT.

Similarly, for the text modeling part, we use $\Mat{E}^2_{word}$ to denote the word embedding matrix of the given problem and $\Mat{E}^2_{pos}$ to denote the corresponding position matrix.
Then the computational formulas for BERT are defined as follows:
\begin{equation}\label{eq:emb2}
\Mat{E}^2 = \Mat{E}^2_{word}+\Mat{E}^2_{pos}\,,
\end{equation}
\begin{equation}\label{eq:bert}
\Mat{H}^2, \Vector{h}^2_{CLS} = \mathrm{BERT}(\Mat{E}^2)\,,
\end{equation}
where $\Mat{H}^2$ is the hidden word representations and $\Vector{h}^2_{CLS}$ is the CLS embedding obtained by BERT.

\subsection{Coupling of Pre-trained Models}
In the above manner, the text part and the code part are separately modeled.
A naive way to fuse these two parts is to just concatenate the hidden representations in the top layer of the model.
However, simple concatenation could not capture the interactions between different modalities in transformer networks of pre-trained models.
In reality, the text of the problem and the code of its corresponding solution are correlated with each other since they are intrinsically paired.
Therefore, it is necessary to model their interactions.

To this end, the proposed model C-BERT utilizes the CLS embeddings to correlate the two parts.
We do not consider concatenating the word and token representations in the inner layers of the transformer networks.
This is because BERT is tailored for general textual words while CodeBERT is suitable for code tokens.
Moreover, correlating the CLS embeddings almost does not increase the computational cost.
We set up two CLS representations in each BERT and CodeBERT, one representing the modal sentence information and one representing the cross-modal interaction information, which is aimed to facilitate the cross-modal interaction.
For the CLS representation corresponding to cross-modal interaction information, its query is equal to the CLS query from another modality.
Specifically, we correlate the two parts by the following fusion function:
\begin{equation}\label{eq:cls-fusion}
\Vector{Q}^{n}_{crossCLS^{2}} = \Vector{Q}^{n}_{CLS^{1}},~~~~~~~    
\Vector{Q}^{n}_{crossCLS^{1}} = \Vector{Q}^{n}_{CLS^{2}},.
\end{equation}
where ${Q}^{n}_{x}$ is the query of x token in layer n.

As such, the interaction modeling is realized by the CLS embeddings when stacking multiple transformer layers.

\subsection{Estimation and Training}\label{subsec:pre-train}
To perform the difficulty level estimation, C-BERT first calculates the overall embeddings for the text and the code.
This is realized by averaging the hidden representations of words and tokens, which are given by:
\begin{equation}\label{eq:overall-embed}
\Vector{h}^1 = \frac{\sum_i \Mat{H}^1_i}{m}\,, \Vector{h}^2 = \frac{\sum_j \Mat{H}^2_j}{n}\,,
\end{equation}
where $\Vector{h}^1$ corresponds to code and $\Vector{h}^1$ corresponds to text.
We do not use the CLS embeddings since they are already used for interaction modeling.
And by our empirical tests, the average embeddings achieve better performance than using the CLS embeddings.

Despite the representations of the text and code, we also consider the feature vector $\Vector{x}_r$ for a POJ problem, which mainly consists of the following features: (1) time limit w.r.t. the operational requirement, (2) space limit w.r.t. the operational requirement, (3) size of input and output, and (4) category label of the problem.
we concatenate the feature vector with the two hidden representations together, i.e., $\Vector{x} = [\Vector{h}^1\oplus\Vector{h}^1\oplus\Vector{x}_r]$.
Consequently, the estimation is computed as follows:
\begin{equation}\label{eq:softmax}
    \Vector{p} = \mathrm{Softmax}\big(\mathrm{MLPs}(\Vector{x})\big)\,,
\end{equation}
\begin{equation}\label{eq:prediction}
    \hat{y} = \arg\max_i \Vector{p}_i\,,
\end{equation}
where $\Vector{p}$ is the predicted probability distribution w.r.t. different difficulty levels and MLP denote multi-layer perceptron.
We then choose the difficulty level with the maximal probability as the estimation $\hat{y}$.

We can adopt the cross-entropy loss to fine-tune the coupled pre-trained language models.
However, the programming language gap between pre-training and testing should be noted.
Actually, the original version of CodeBERT is pre-trained by the codes written by several languages (e.g., Python, Java) but without C and C++.
However, C and C++ are largely used in POJ platforms for code practice.
To mitigate this gap, we perform additional pre-training of CodeBERT on the codes from the POJ platforms that we used in the experiments.

\section{EXPERIMENTS}
In this section, we first show the experimental settings and then analyze the experimental results.

\subsection{Experimental Setup}
\textbf{Datasets}. We adopt two public datasets released in Description2code\footnote{https://github.com/ethancaballero/description2code} to build the datasets for the studied task in this paper.
The first raw dataset is collected from the Codeforces platform and the second raw dataset is from CodeChef.
Hence, we use Codeforces and CodeChef to name the two datasets.
Since the released datasets have no difficulty labels, we crawl them from the corresponding platforms.
To obtain the example solutions, we further collect the source codes submitted by one active programmer.
We adopt some text processing techniques, such as removing abnormal samples, removing stop words, etc.
According to the characteristics of POJ problems and solutions, we also standardize the symbols of problem statements and remove the comments of the codes.
Finally, the statistics of the two built datasets are shown in Table~\ref{tbl:statistics}.
As can be seen, the number of difficulty levels is 3 for Codeforces and 5 for CodeChef.

\begin{table}[!t]
  \centering
  \caption{Statistics of the datasets.}\label{tbl:statistics}
  \vspace{-1em}
\begin{tabular}{c|cc}
\hline
\textbf{Dataset} & \textbf{\#problem} & \textbf{\#class} \\ \hline
Codeforces & 5302 & 3 \\
CodeChef & 1223 & 5 \\\hline
\end{tabular}
\end{table}

\noindent\textbf{Evaluation settings}.
We employ 5-fold cross-validation to conduct performance comparisons among the proposed model and baselines.
All the hyperparameters of the adopted methods are tuned on one fold and the performance is tested on the left four folds.
This makes the comparison more reliable.
Since this task is actually a multi-class classification problem, we adopt the accuracy rate, F1 score, and AUC score for evaluation. For the accuracy rate and F1 score, we use the macro mode. And for AUC, we use the OVR mode.

\noindent\textbf{Baselines}.
We choose the following methods as the baselines.

\textbf{XGBoost}~\cite{chen2016xgboost} is a commonly-used strong multi-class classification model.
Despite the explicit features used in C-BERT, we additionally extract text and code features for XGBoost, including the features w.r.t the problem statement (e.g., text length, number of numerical values in the statement), and the features w.r.t. the code (e.g., length of the code, keywords, number of loops, number of nodes and edges in the Abstract Syntax Tree (AST), Data Flow Graph (DFG),  Control Flow
Graph (CFG), and Program Dependency Graph (PDG)).

\textbf{BERT} is the pre-trained language model for the text part of POJ problems.
To make it suitable for the task, we also fine-tune it in a supervised fashion.

\textbf{BERT+} is a simple extension of BERT to combine the text modality and the code modality as input.
It further concatenates the explicit feature vector as ours.


\textbf{CodeBERT+} is the pre-trained model for the code modality.
Similar to BERT+, it combines the two modalities and incorporates features as well.

\textbf{GraphCodeBERT}~\cite{guo2021graphcodebert} is a variant of CodeBERT that exploits DFG of the source code. 

\textbf{Devign}~\cite{zhou2019devign} is a standard graph neural network-based model that is tailored for code vulnerability detection.
Here we utilize it for difficulty level estimation by only modifying the output layers of the model.

\noindent\textbf{Implemention details.}
The version of BERT used throughout this paper is the base one with a parameter size of 110M.
The version of CodeBERT is also the base one with a parameter size of 125M.
The length settings (512 by default) of BERT and CodeBERT are tuned based on the performance and memory space limitation.
The dimension of Devign is tuned to be 128.
The number of trees for XGBoost is 160.
We run all the experiments on a Linux server with a GTX2080TI GPU card.
The optimization algorithms are based on gradient descent methods, with tuned learning rates.

\begin{table}[!t]
  \centering
  \caption{Performance comparison on the two datasets.}\label{tbl:comparison}
  \vspace{-1em}
\resizebox{1.\linewidth}{!}{
\begin{tabular}{l|ccc|ccc}
\hline
            \multirow{2}{*}{Methods}           & \multicolumn{3}{c|}{CodeForces} & \multicolumn{3}{c}{CodeChef} \\ \cline{2-7}
                      & AUC & ACC & F1 & AUC & ACC & F1 \\ \hline
XGBoost & 79.83 & 67.52 & 67.37 & 67.56 & 43.25 & 39.54 \\
BERT & 70.48 & 54.19 & 53.07 & 64.52 & 40.21 & 34.23 \\
BERT+ & 84.62    & 67.82    & 68.85    & 72.87    & 46.68   & 39.61\\
CodeBERT+ & 82.93    & 65.39    & 66.05    & 70.29    & 45.45   & 38.81\\
GraphCodeBERT & 82.35 & 66.74 & 66.95 & 68.99 & 42.99 & 38.56 \\
Devign & 79.87 & 64.56 & 65.01 & 67.05 & 41.13 & 36.56 \\
\hline
C-BERT & \textbf{87.25} & \textbf{70.49} & \multicolumn{1}{c|}{\textbf{71.01}} & \textbf{76.04} & \textbf {49.23} & \textbf{42.15} \\

\hline
\end{tabular}
}
\end{table}

\subsection{Model Comparison}
We compare all the adopted models on the two datasets.
The results are shown in Table~\ref{tbl:comparison}, from which we have the following key observations:
\begin{itemize}[leftmargin=*]
\item[$\diamond$] BERT performs not as well as other models.
This indicates that only modeling the text modality is challenging for the difficulty level estimation task.
It is intuitive since the statement of a POJ problem usually involves some background stories which are not directly related to the programming difficulty.

\item[$\diamond$] Although XGBoost is a strong competitor in many classification tasks and we design many hand-crafted features for this model, it behaves only as well as GraphCodeBERT which uses the single code modality.
Moreover, XGBoost is significantly inferior to BERT+.
These comparisons reveal the benefits of considering pre-trained models for this task.

\item[$\diamond$] BERT+ outperforms BERT on the two datasets consistently.
This is attributed to the fact that BERT+ additionally uses the source code as model input and further incorporates the features w.r.t. operational requirement.

\item[$\diamond$] The graph neural network-based model Devign uses different graph structure information for code modeling, including not only DFG used in GraphCodeBERT, but also AST and CFG.
Nevertheless, it is still not as well as GraphCodeBERT.
This again consolidates the necessity of utilizing pre-trained models for the novel task formulated in this paper.

\item[$\diamond$] Finally, the proposed model C-BERT achieves superior performance among all the models.
This might be attributed to the characteristics of the model, including using domain-specific pre-trained models for different modalities and the coupling of the two pre-trained models.

\end{itemize}

\subsection{Ablation Study}
In this section, we further conduct an in-depth analysis to show whether the main components of C-BERT have positive contributions to the final performance.
In particular, we design the following variants of the full model to achieve an ablation study. 

\textbf{w/o Text} means removing the textual part of C-BERT.
As a consequence, the branch w.r.t. BERT is removed, but the features and CodeBERT are kept within the variant.

\textbf{w/o Code} means removing the branch w.r.t. CodeBERT and thus the features and BERT are kept.

\textbf{w/o Feature} modifies the concatenation operation $\Vector{x} = [\Vector{h}^1\oplus\Vector{h}^1\oplus\Vector{x}_r]$ by $\Vector{x} = [\Vector{h}^1\oplus\Vector{h}^1]$.
Hence the information w.r.t. the operational requirement is deleted.

\textbf{w/o Coupling} denotes not considering the coupling of the two pre-trained models.
Therefore, Equation~\ref{eq:cls-fusion} is not involved in the computational procedure of C-BERT.

\begin{table}[!t]
  \centering
  \caption{Results of ablation study for C-BERT.}\label{tbl:ablation}
  \vspace{-1em}
\resizebox{1.\linewidth}{!}{
\begin{tabular}{l|ccc|ccc}
\hline
            \multirow{2}{*}{Methods}              & \multicolumn{3}{c|}{CodeForces} & \multicolumn{3}{c}{CodeChef} \\ \cline{2-7} 
                      & AUC      & ACC      & F1       & AUC      & ACC     & F1      \\ \hline

\multicolumn{1}{l|}{C-BERT}                 & \textbf{87.25}             & \textbf{70.49}             & \multicolumn{1}{l|}{\textbf{71.01}} & \textbf{76.04}             & \textbf{49.23}             & \textbf{42.43}             \\\hline
- w/o Text & 84.71    & 68.01    & 68.53    & 70.25    & 44.45   & 39.51  \\
- w/o Code & 84.07    & 67.23    & 68.02    & 71.54    & 44.67   & 37.69   \\
- w/o Feature & 86.25    & 69.32    & 69.99    & 73.90    & 47.23   & 41.01  \\
- w/o Coupling & 85.42    & 68.14    & 69.08    & 73.12    & 47.68   & 39.87   \\ \hline

\end{tabular}
}
\end{table}

Table~\ref{tbl:ablation} presents the performance results of C-BERT and its four variants.
Based on the results, we find that:

\begin{itemize}[leftmargin=*]
\item All the four variants are notably worse than the full model on the two datasets.
This meets the expectation that using BERT for problem statement modeling, using CodeBERT for code modeling, incorporating features into difficulty estimation, and coupling pre-trained models all have positive contributions.

\item The variant of ``w/o Code'' suffers from the largest performance degradation in most cases. 
This indicates the part of code modeling is critical for the difficulty level estimation task, which is consistent with the previous observation.

\end{itemize}

\section{Conclusion}
In this paper, we formulate a novel task of difficulty level estimation of programming problems.
This task facilitates the programming practice of students in POJ platforms and provides the multi-modal perspective involving text and code for research study
To solve this task, we propose to leverage pre-trained models and combine them in a unified model named C-BERT.
C-BERT couples the two models by correlating the CLS embeddings of each model.
We build two real POJ datasets for task evaluation.
The experiments demonstrate that C-BERT outperforms several candidate strong methods and validate the effectiveness of the main components within the model.


\bibliographystyle{ACM-Reference-Format}
\bibliography{acmart}

\end{document}